\documentclass[]{tMOP2eN}
\usepackage{epsfig}
\usepackage{epstopdf}
\usepackage{dsfont}
\usepackage{soul}
\usepackage{color}
\usepackage[colorlinks]{hyperref}
\hypersetup{
	bookmarksnumbered,
	pdfstartview={FitH},
	citecolor={darkgreen},
	linkcolor={darkred},
	urlcolor={darkblue},
	pdfpagemode={UseOutlines}}
\definecolor{darkgreen}{RGB}{50,190,50}
\definecolor{darkblue}{RGB}{0,0,190}
\definecolor{darkred}{RGB}{238,0,0}
\DeclareMathOperator{\atanh}{atanh}
\DeclareMathOperator{\diag}{diag}
\DeclareMathOperator{\arsinh}{arsinh}
\begin{document}
\doi{10.1080/09500340.2012.712725}
 \issn{1362-3044}
\issnp{0950-0340} 
\jyear{2012} \jmonth{September}

\title{Entanglement generation in relativistic quantum fields}
%
%
%
%
\author{
Nicolai Friis$^{\dagger}$\thanks{$^\dagger$Corresponding author. Email: pmxnf@nottingham.ac.uk \vspace{6pt}} and
Ivette Fuentes$^{\ddagger}$\thanks{$^\ddagger$Previously known as Fuentes-Guridi and Fuentes-Schuller. \vspace{6pt}}\\ 
{\em{School of Mathematical Sciences, University of Nottingham, University Park, Nottingham NG7 2RD, United Kingdom}};
\\  \vspace{6pt}\received{September 2012} }
\maketitle
\begin{abstract}
We present a general, analytic recipe to compute the entanglement that is
generated between arbitrary, discrete modes of bosonic quantum fields by
Bogoliubov transformations. Our setup allows the complete characterization
of the quantum correlations in all Gaussian field states. Additionally, it
holds for all Bogoliubov transformations. These are commonly applied in
quantum optics for the description of squeezing operations, relate the mode
decompositions of observers in different regions of curved spacetimes, and
describe observers moving along non-stationary trajectories. We focus on a
quantum optical example in a cavity quantum electrodynamics setting:
an uncharged scalar field within a cavity provides a model for an optical
resonator, in which entanglement is created by non-uniform acceleration.
We show that the amount of generated entanglement can be magnified by initial
single-mode squeezing, for which we provide an explicit formula.
Applications to quantum fields in
curved spacetimes, such as an expanding universe, are discussed.
\end{abstract}
\bigskip
%
%

\section{\label{sec:intro}Introduction}

Over the past decade the discipline of relativistic quantum information has received
much attention (see Ref.~\cite{Czachor1997,PeresScudoTerno2002,GingrichAdami2002,AhnLeeMoonHwang2003,
JordanShajiSudarshan2007,FriisBertlmannHuberHiesmayr2010,HuberFriisGabrielSpenglerHiesmayretal2011}
for a selection and Refs.~\cite{PeresTerno2004,RalphDownes2012} for reviews). Its aim is the study of
the resources and tasks of quantum information science in the context of relativity. In particular,
finding suitable ways to store and process information is a main goal.

It has therefore been a central focus of previous efforts to identify the \emph{degradation} effects
due to relativistic, accelerated motion~\cite{Fuentes-SchullerMann2005,
AlsingFuentes-SchullerMannTessier2006,AdessoFuentes-SchullerEricsson2007,
BruschiLoukoMartin-MartinezDraganFuentes2010,Martin-MartinezFuentes2011,
FriisKoehlerMartinMartinezBertlmann2011,MonteroLeonMartinMartinez2011} and spacetime
curvature~\cite{AdessoFuentes-Schuller2009,Martin-MartinezGarayLeon2010a}. At the centre of
these deteriorations lie two ingredients: the Bogoliubov transformation of the mode
operators and the presence of a horizon.
While the Bogoliubov transformations are responsible for the creation and shift of excitations,
the horizon causes some of these excitations to be inaccessible to the observer. This usually
leads to a loss of information and, consequently, a degradation of the quantum correlations.
A rare exception of a situation where entanglement is \emph{generated} by acceleration can
be found in Ref.~\cite{MonteroMartin-Martinez2011}.

In Ref.~\cite{DownesFuentesRalph2011} cavities were proposed as a suitable way of storing and
processing information in the relativistic context. Cavities, represented by appropriate
boundary conditions, can be uniformly or non-uniformly accelerated without
creating a horizon. Nevertheless, the absence of an event horizon does not guarantee complete
access to all modes of a quantum field. In particular, it was shown in
Refs.~\cite{BruschiFuentesLouko2012,FriisLeeBruschiLouko2012} how the entanglement between
a non-uniformly moving cavity and an inertial reference cavity is degraded when only
particular modes are being considered. In this case the information loss can be
prevented by appropriately timing the periods of uniform acceleration and inertial motion.

However, the role of the Bogoliubov transformations lies foremost in the \emph{generation of
entanglement}, an effect that can be found for a wide variety of situations. In quantum optics
they are used for the description of single-mode as well as multi-mode squeezing operations,
see, e.g., Ref.~\cite{GarrisonChiao:QuantumOptics}. Entanglement is also created between modes
of quantum fields in curved spacetimes
~\cite{BallFuentes-SchullerSchuller2006,Martin-MartinezGarayLeon2010b}. And, recently,
entanglement generation has been demonstrated and studied quantitatively for non-uniform cavity
motion in flat Minkowski spacetime in
Refs.~\cite{FriisBruschiLoukoFuentes2012,BruschiDraganLeeFuentesLouko2012}.

Here we present a framework that generalises and significantly simplifies the previous approaches
on entanglement generation in relativistic quantum information. We employ continuous variable
techniques~\cite{AdessoIlluminati2005} that allow us to investigate the quantum correlations that
are created by arbitrary Bogoliubov transformations of any discrete number of modes. The
transformed state of these modes is described by its covariance matrix, which completely
determines the entanglement of the system. Results for any given Bogoliubov transformation can
thus be obtained analytically in principle, if the transformation coefficient are known.

We apply our method to a quantum optical setting: an optical cavity that is moving
non-uniformly is modelled by confining an uncharged scalar field by appropriate boundary
conditions~\cite{BruschiFuentesLouko2012}. The cavity is following a trajectory that consists of
segments of uniform acceleration and inertial motion. We work in a perturbative regime
where the products of the cavity's width and the individual accelerations in all segments are small,
but all calculations can be carried out analytically.
The Bogoliubov coefficients can be expressed as a Maclaurin series in $h$, a parameter which
represents the product of the acceleration at the centre of the cavity and the cavity's width
$\delta$. The creation of entanglement for initially uncorrelated, single-mode squeezed states
of the field is quantified in terms of the expansion coefficients of the power series. We find that
the entanglement generation in this fully relativistic setting can be enhanced by initial
single-mode squeezing. This provides an additional controllable parameter for possible experimental
setups to enhance the visibility of the effect.

We further emphasize the applicability of our approach beyond cavity quantum electrodynamics by
verifying the results obtained for an expanding universe in Ref.~\cite{BallFuentes-SchullerSchuller2006}
as a curved spacetime examples.

This article is structured as follows. In Sec.~\ref{sec:bogo and gaussian states} we establish the
basic description of Bogoliubov transformations for arbitrary Gaussian initial states in terms of
the corresponding symplectic transformation of the covariance matrix. In
Sec.~\ref{sec:bogo bogo boxes} we revise the description of non-uniformly moving cavities from
Refs.~\cite{BruschiFuentesLouko2012,FriisLeeBruschiLouko2012} and, subsequently, apply the formalism
of Sec.~\ref{sec:bogo and gaussian states} to study the entanglement creation by motion in this
scenario.

\section{\label{sec:bogo and gaussian states}Bogoliubov transformation of Gaussian states}

It is the aim of this section to present a general machinery that allows a simple characterisation of
the mode entanglement of bosonic quantum fields. In particular we aim to construct a framework that
works for any specified Bogoliubov transformation.

To this end, let us consider an arbitrary discrete set of mode functions $\left\{\phi_{n}\,|\,n=1,2,3,\ldots\right\}$
of a bosonic quantum field, let us assume a scalar field for simplicity, with associated annihilation
and creation operators, $a_{n}$ and $a_{n}^{\dagger}$, respectively. The field operators satisfy the
canonical commutation relations $[a_{m},a_{n}]=[a_{m}^{\dagger},a_{n}^{\dagger}]=0$ and
$[a_{m},a_{n}^{\dagger}]=\delta_{mn}$.
The functions $\phi_{n}$, which we take to be the solutions to a (relativistic) field equation, e.g.,
the Klein-Gordon equation, form a complete set of orthonormal modes with respect to a chosen scalar
product, i.e., $(\phi_{m},\phi_{n})=\delta_{mn}$, see, e.g., Ref.~\cite{BirrellDavies:QFbook}.
In the usual Fock representation the vacuum state is annihilated by all $a_{n}$, i.e.,
$a_{n}\,|\,0\,\rangle=0,\,\forall\,n$, while particle states are created by the (repeated) action of the
creation operators $a_{n}^{\dagger}$.

We can then perform a \emph{Bogoliubov transformation} that relates our initial modes $\phi_{n}$ and the
associated operators $a_{n}$ and $a_{n}^{\dagger}$ to another (complete, orthonormal) set of modes
\mbox{$\{\tilde{\phi}_{n}\,|\,n=1,2,3,\ldots\}$,} with operators $\tilde{a}_{n}$ and
$\tilde{a}_{n}^{\dagger}$, respectively, i.e.,
\begin{equation}
    \tilde{a}_{m}\,=\,\sum\limits_{n}\,\bigl(\,\alpha^{*}_{mn}\,a_{n}\,-\,\beta^{*}_{mn}\,a^{\dagger}_{n}\,\bigr)\,.
    \label{eq:general bogo operators}
\end{equation}
Here we have used the conventions of Ref.~\cite{BirrellDavies:QFbook}. The asterisk denotes complex
conjugation and $\alpha_{mn}=(\tilde{\phi}_{m},\phi_{n})$ and $\beta_{mn}=-(\tilde{\phi}_{n},\phi^{*}_{m})$
are the coefficients of the Bogoliubov transformation.

This transformation can represent a physical operation in a laboratory, e.g., a squeezing operation in an
optical setup, or a change of potential in a harmonic oscillator chain of trapped atoms. Alternatively,
the transformation may describe a change of observer in a (curved) spacetime, e.g., a black hole spacetime
or a cosmological model, or relate the modes at
different times of a single observer that is moving non-uniformly.
For a continuous spectrum of solutions the sum in Eq.~(\ref{eq:general bogo operators}) has to be replaced
by an appropriate integral, but we will specialise to discrete sets of modes in the following.

Analogously to before, the vacuum state $|\,\tilde{0}\,\rangle$ of the new mode decomposition is defined by
the property $\tilde{a}_{n}\,|\,\tilde{0}\,\rangle=0,\,\forall\,n$. However, an important feature of the
Bogoliubov transformations is the non-equivalence of the corresponding vacua, i.e.,
$a_{m}\,|\,\tilde{0}\,\rangle=0$ only if $\beta_{mn}=0,\,\forall\,n$. Consequently, the vacua generally have
a nontrivial relationship, see, e.g., Ref.~\cite{FabbriNavarro-Salas2005}.

To study the quantum correlations between particular modes two more steps are required. The complete Fock
state basis of the original modes has to be expressed in terms of the Fock states of the transformed modes.
Subsequently, the subset of modes that is not considered must be traced over. While the first step can
generally be accomplished, the tracing procedures are usually cumbersome.

In the following we demonstrate how these computations can be simplified significantly. Let us switch to
the covariance matrix formalism discussed in Ref.~\cite{AdessoIlluminati2005}. This means, instead of the complete Fock space description of the quantum state, we consider only the matrix $\sigma$ with components
\begin{align}
    \sigma_{ij}\,=\,\left\langle\,\mathrm{X}_{i}\mathrm{X}_{j}\,+\,\mathrm{X}_{j}\mathrm{X}_{i}\,\right\rangle\,
	   -\,2\,\left\langle\,\mathrm{X}_{i}\,\right\rangle\left\langle\,\mathrm{X}_{j}\,\right\rangle\,.
    \label{eq:covariance matrix}
\end{align}
The quadrature operators $\mathrm{X}_{i}$ are chosen to be the generalised positions and momenta, i.e.,
$\,\mathrm{X}_{(2n-1)}=\tfrac{1}{\sqrt{2}}(a_{n}+a_{n}^{\dagger})\,$ and
$\,\mathrm{X}_{(2n)}=\tfrac{-i}{\sqrt{2}}(a_{n}-a_{n}^{\dagger})\,$,
where the index $n=1,2,3,\ldots$ labels the modes, and $\langle\,\mathcal{O}\,\rangle$ denotes the expectation value of the operator $\mathcal{O}$. The real, symmetric covariance matrix $\sigma$ together with the vector of first moments $\mathrm{X}_{i}$ completely characterises all Gaussian states, i.e., states that can be described by quasi-probability distributions of Gaussian shape in phase space, see, e.g., Ref.~\cite{AdessoIlluminati2005}. However, the covariance matrix itself is a sufficient description of all properties pertaining to entanglement. While these continuous variable tools have been frequently employed in quantum optics settings, their application to quantum field theory and relativistic quantum information was previously explored only in limited situations~\cite{AdessoFuentes-SchullerEricsson2007,AdessoFuentes-Schuller2009}.

Unitary transformations in the Fock space are represented by symplectic transformations in phase space. A transformation $S$ is called symplectic, if it leaves the symplectic form $\Omega$ invariant, i.e., $S\,\Omega\,S^{T}=\Omega$, where $[X_{i},X_{j}]=i\Omega_{ij}$ and $(S_{ij})^{T}=(S_{ji})$.
From Eq.~(\ref{eq:general bogo operators}) it is straightforward to find the expression for the
symplectic representation $S$ of a Bogoliubov transformation in terms of its
general coefficients $\alpha_{mn}$ and $\beta_{mn}$. The matrix~$S$ is decomposed into $2\times2$ blocks $\mathcal{M}_{mn}$ as
\begin{align}
    S\,=\,\begin{pmatrix}
        \mathcal{M}_{11}  &   \mathcal{M}_{12}  &   \mathcal{M}_{13}   &   \ldots  \\
        \mathcal{M}_{21}  &   \mathcal{M}_{22}  &   \mathcal{M}_{23}   &   \ldots  \\
        \mathcal{M}_{31}  &   \mathcal{M}_{32}  &   \mathcal{M}_{33}   &   \ldots  \\
        \vdots            &   \vdots            &   \vdots             &   \ddots  \\
    \end{pmatrix}\,,
    \label{eq:Gaussian n-mode Bogo transformation}
\end{align}
where the sub-blocks are given by
\begin{align}
    \mathcal{M}_{mn}\,=\,
        \begin{pmatrix}
            \Re(\alpha_{mn}\,-\,\beta_{mn})     &   \Im(\alpha_{mn}\,+\,\beta_{mn}) \\[1.5mm]
            -\Im(\alpha_{mn}\,-\,\beta_{mn})    &   \Re(\alpha_{mn}\,+\,\beta_{mn}) \\
        \end{pmatrix}\,.
    \label{eq:M Bogo matrix}
\end{align}
Here $\Re(z)$ and $\Im(z)$ denote the real part and imaginary part of the complex number $z$ respectively. The
transformed covariance matrix $\tilde{\sigma}$ is then simply obtained as
\begin{align}
    \tilde{\sigma}\,=\,S\,\sigma\,S^{T}\,.
    \label{eq:transformed covariance matrix}
\end{align}
The crucial technical simplification lies in the ensuing step.
The partial trace over any subset of modes $\mathcal{E}\subset\{n=1,2,3,\ldots\}$ can be computed trivially by
eliminating all rows and columns corresponding to modes $m\in\mathcal{E}$. This technique applies for any initial state $\sigma$. However, if we restrict ourselves to Gaussian states, the unitarity of the transformation ensures that $\tilde{\sigma}$ remains Gaussian, which allows us to use the specifically useful tools for the quantification of Gaussian entanglement~\cite{AdessoSerafiniIlluminati2006}. In particular, we are interested in the generation of entanglement from initially uncorrelated states. In this case the covariance matrix $\sigma$ has a block diagonal structure, $\sigma=\diag\{\psi_{n}\}$, where the $2\times2$ blocks $\psi_{n}$ are locally equivalent to single-mode squeezed states with squeezing parameters $s_{n}$, i.e.,
\begin{align}
    \psi_{n}\,=\,
        \begin{pmatrix}
            e^{s_{n}}     &   0 \\
            0    &   e^{-s_{n}} \\
        \end{pmatrix}\,.
    \label{eq:single mode squeezing}
\end{align}
Let us now consider the transformed state of two modes, labelled $k$ and $k^{\prime}$ respectively. In other words, \mbox{$\mathcal{E}=\{n=1,2,3,\ldots|n\neq k,k^{\prime}\}$} is the set of all modes except the chosen pair $k$ and $k^{\prime}$. The transformed covariance matrix $\tilde{\sigma}_{kk^{\prime}}$ of these two modes is then given by
\begin{align}
    \tilde{\sigma}_{kk^{\prime}}\,=\,\begin{pmatrix}
        C_{kk}\ \  &   C_{kk^{\prime}}  \\[1mm]
        C_{k^{\prime}k}\ \  &   C_{k^{\prime}k^{\prime}}  \\
    \end{pmatrix}\,,
    \label{eq:transformed two mode covariance matrix}
\end{align}
where $C_{ij}=\sum
_{n}\mathcal{M}_{in}\,\psi_{n}\,\mathcal{M}_{jn}^{T}$.

Let us now turn to the quantification of the entanglement that is generated between the modes $k$
and~$k^{\prime}$.
The quantity that encodes all the information about the entanglement of symmetric two-mode Gaussian
states is the smallest symplectic eigenvalue $\widehat{\nu}_{-}$ of
$\,\widehat{\sigma}_{kk^{\prime}}=T_{k^{\prime}}\,\tilde{\sigma}_{kk^{\prime}}\,T_{k^{\prime}}\,$,
where $T_{k^{\prime}}=\diag\{1,1,1,-1\}$ represents the partial transposition of mode~$k^{\prime}$,
see Ref.~\cite{AdessoIlluminati2005}.
For this class of states the Gaussian measures of entanglement, such as the (logarithmic) negativity
or the Gaussian entanglement of formation, are monotonously decreasing functions of
$\widehat{\nu}_{-}$.
However, the two-mode Gaussian state $\tilde{\sigma}_{kk^{\prime}}$ of
Eq.~(\ref{eq:transformed two mode covariance matrix}) is generally not symmetric with respect to the
interchange of $k$ and $k^{\prime}$. Consequently, different entanglement measures can exhibit
different ordering of two entangled states.

With this in mind we specialise to the use of the negativity $\mathcal{N}$ as an entanglement measure.
It has the advantage of being easily computable, i.e., the negativity of $\sigma_{kk^{\prime}}$ is
given in terms of the smallest symplectic eigenvalue $\widehat{\nu}_{-}$ of
$\widehat{\sigma}_{kk^{\prime}}$ by the simple formula
\begin{align}
    \mathcal{N}=\max\{0,(1-\widehat{\nu}_{-})/2\widehat{\nu}_{-}\}\,.
    \label{eq:negativity def}
\end{align}
Furthermore, it can be easily
compared to results obtained for non-Gaussian states in Ref.~\cite{FriisBruschiLoukoFuentes2012}.
The smallest symplectic eigenvalue is obtained by diagonalising the matrix
$\widehat{\sigma}_{kk^{\prime}}$ by a symplectic operation $D$, such that
$\,D\,\widehat{\sigma}_{kk^{\prime}}\,D^{T}=
\diag\{-\widehat{\nu}_{-},\widehat{\nu}_{-},-\widehat{\nu}_{+},\widehat{\nu}_{+}\}\,$,
where $0\leq\widehat{\nu}_{-}\leq\widehat{\nu}_{+}$.
The quantities $\pm\widehat{\nu}_{\pm}$ can be computed as the eigenvalues of
$i\Omega\widehat{\sigma}_{kk^{\prime}}$ in a straightforward way.
For $0\leq\widehat{\nu}_{-}<1$ the state $\tilde{\sigma}_{kk^{\prime}}$ is entangled.


Other well known entanglement measures for Gaussian states, e.g., Gaussian entanglement of formation, can
also be calculated straightforwardly for two-mode states in principle~\cite{AdessoIlluminati2005}. The
transformed state will generally be a mixed state but all calculations can be done analytically, provided
that the Bogoliubov coefficients are given and the infinite sums in
Eq.~(\ref{eq:transformed two mode covariance matrix}) are convergent.
In Sec.~\ref{sec:bogo bogo boxes} we discuss an example of a discrete spectrum for which the
Bogoliubov coefficients are given as a perturbative expansion and the sums
converge~\cite{BruschiFuentesLouko2012,FriisLeeBruschiLouko2012}. We compute the entanglement generated from
non-uniform cavity motion for particular initial states of interest. For continuous spectra, however, the
corresponding integrals are often known to be divergent, see, e.g., Ref.~\cite{Schuetzhold2001}, thus rendering
any conclusive statements about the entanglement generated between the modes fruitless, unless the coefficients
have a simple structure.

Such a situation presents itself in the case of a charged, scalar field in an expanding universe, which was discussed
in Ref.~\cite{BallFuentes-SchullerSchuller2006}. There the spectrum of the quantum field is continuous, but the
Bogoliubov transformation between the asymptotically flat remote past and future couple only modes of opposite
momenta. In this fashion, the transformation provides an effective discretization and we can reproduce the results
of Ref.~\cite{BallFuentes-SchullerSchuller2006} with our methods.

\section{\label{sec:bogo bogo boxes}Non-uniform cavity motion}

An example for a discrete, bosonic spectrum is obtained by confining a scalar quantum field
to a cavity by appropriate Dirichlet boundary conditions in $(1+1)$ dimensions. As proposed in
Ref.~\cite{BruschiFuentesLouko2012} the rigid cavity can follow a worldline that is composed of
segments of inertial motion and uniform acceleration. This non-uniform motion generates
non-trivial Bogoliubov coefficients, which result in a generation of entanglement between the
modes inside one cavity~\cite{FriisBruschiLoukoFuentes2012} and, consequently, lead to a
degradation of initial entanglement between modes in different cavities~\cite{BruschiFuentesLouko2012}.

For a quantitative description the Bogoliubov transformations are expanded as a Maclaurin series in the
small, dimensionless parameter $h$, i.e.,
\begin{subequations}
\label{eq:alphas and betas small h expansion}
\begin{align}
\alpha  &=  \alpha^{(0)} + \alpha^{(1)} + \alpha^{(2)} + O(h^{3})\,,
\label{eq: alphas small h expansion}\\
\beta   &=  \beta^{(1)} + \beta^{(2)} + O(h^{3})\,,
\label{eq:betas small h expansion}
\end{align}
\end{subequations}
where the superscripts $^{(n)}$ denote quantities proportional to $h^{n}$. The parameter $h$ is the product of
the cavity's length in its instantaneous rest frame and the proper acceleration at the centre of the cavity.
Here we use units such that Planck's constant and the speed of light are dimensionless constants, $\hbar=c=1$,
and the entity $O(x)/x$ is bounded as $x$ goes to $0$. Furthermore, the coefficient $\alpha^{(0)}$ must include
the phases of the free time evolution in the uniformly accelerated and inertial segments, while it reduces to
the identity for vanishing accelerations. We therefore have $\alpha^{(0)}_{mn}=\delta_{mn}G_{m}$, where $G_{m}$
is a mode-dependent phase factor of unit magnitude, i.e., $|G_{m}|=1$. Additionally, we find that the linear
corrections vanish on the diagonal, i.e., $\alpha^{(1)}_{nn}=\beta^{(1)}_{nn}=0$.

\begin{figure}
\begin{center}
\subfigure[]{
\resizebox*{0.48\textwidth}{!}{\includegraphics{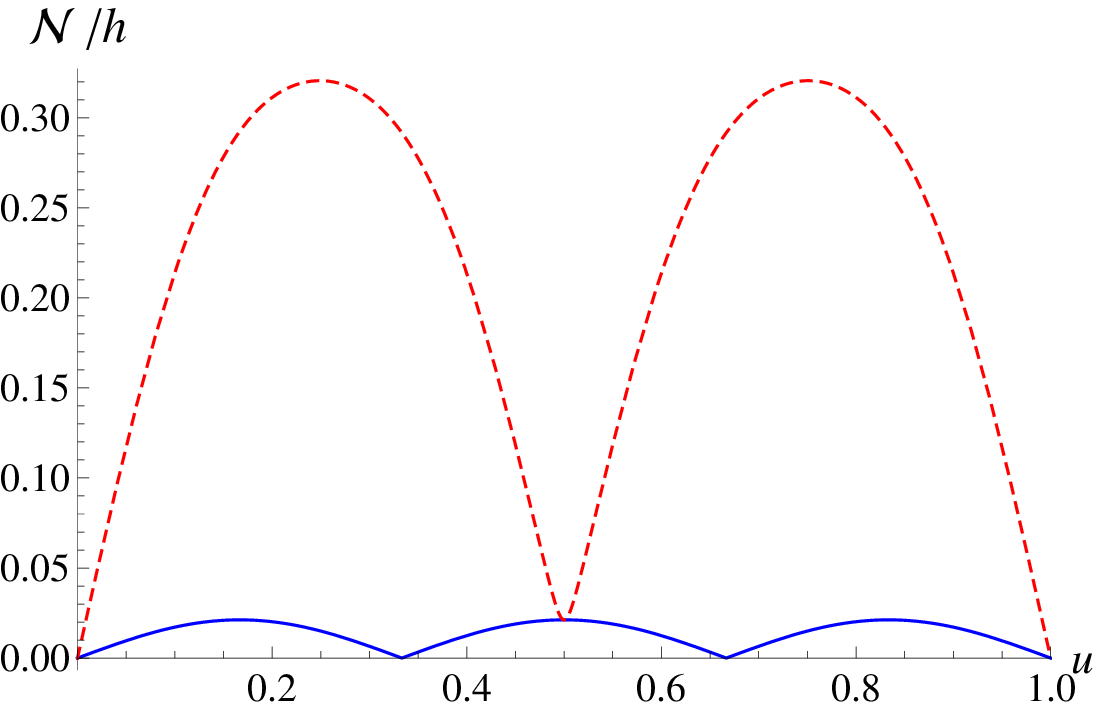}}}%
\subfigure[]{
\resizebox*{0.48\textwidth}{!}{\includegraphics{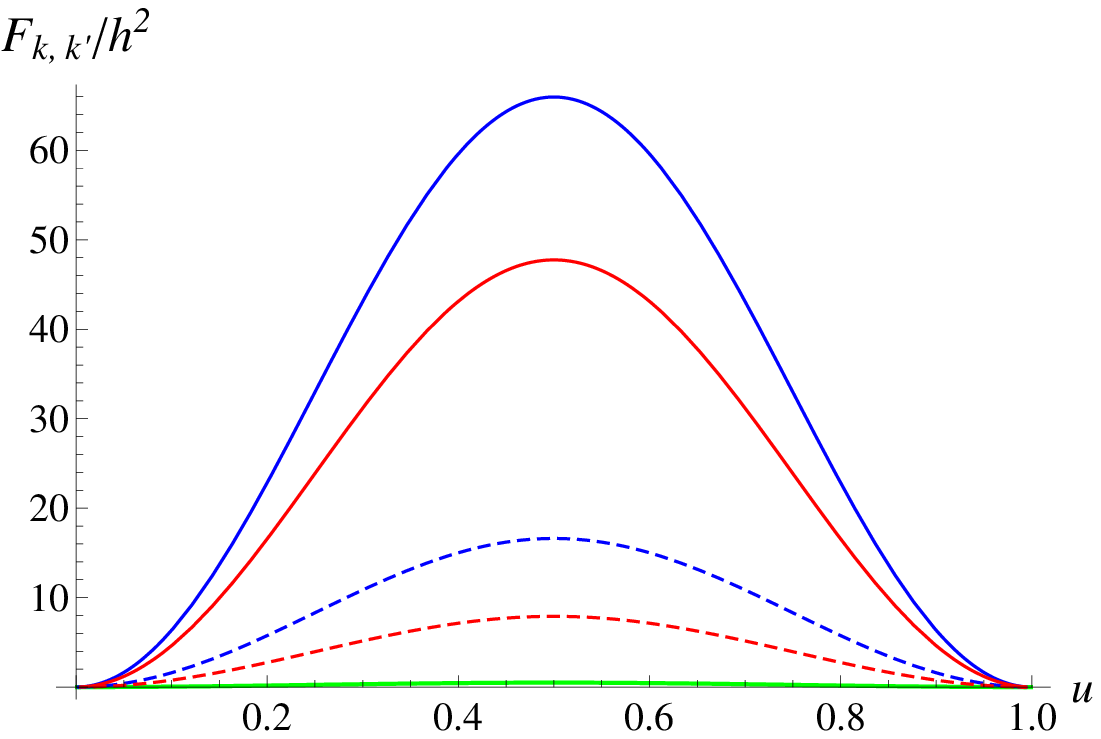}}}%
\caption{The leading order contribution $\mathcal{N}/h$ to the negativity, see
Eq.~(\ref{eq:single mode squeezed negativity}), is shown in
Fig.~\ref{fig:BBB illustration}(a) for a massless scalar field confined to a moving cavity
of width $\delta$. The cavity's trajectory is inertial outside a single segment of uniform,
linear acceleration $h/\delta$ of duration $\tau$, as experienced at the
centre of the cavity. The Bogoliubov coefficients for this travel scenario can be found in
Ref.~\cite{BruschiFuentesLouko2012}. Plots are shown for an initial state of the chosen modes
$(k,k^{\prime})=(1,2)$ that is symmetrically single-mode squeezed with $s=1$ (dashed, red) or
in the (displaced) vacuum, $s=0$ (solid, blue).
Fig.~\ref{fig:BBB illustration}(b) illustrates the behaviour of the quantities
$F_{k,k^{\prime}}/h^{2}=(f^{\alpha}_{k\lnot k^{\prime}} + f^{\alpha}_{k^{\prime}\lnot k})/h^{2}$
for increasing mode numbers. The same field and travel scenario as in
Fig.~\ref{fig:BBB illustration}(a) are used, but the mode numbers are varied. From bottom to
top the curves show $(k,k^{\prime})=(1,2)$ (solid, green), $(k,k^{\prime})=(10,11)$ (dashed, red),
$(k,k^{\prime})=(1,10)$ (dashed, blue), $(k,k^{\prime})=(20,21)$ (solid, red) and
$(k,k^{\prime})=(1,20)$ (solid, blue). All curves are plotted as functions of
the temporal parameter $u:=h\tau/[4\delta\atanh(h/2)]$.}
\label{fig:BBB illustration}
\end{center}
\end{figure}

The perturbative calculations require a closer inspection of the techniques for the calculation of the symplectic
eigenvalues $\widehat{\nu}_{-}$ from Sec.~\ref{sec:bogo and gaussian states}. Let us assume that the initial state
$\sigma_{kk^{\prime}}$ is transformed by the Bogoliubov transformation according to
\begin{align}
    \tilde{\sigma}_{kk^{\prime}}    &=  \sigma_{kk^{\prime}}    +   \sigma^{c}_{kk^{\prime}}\,,
    \label{eq:perturbed covariance matrix}
\end{align}
where$\sigma^{c}_{kk^{\prime}}$ is a small correction to the initial state. For uncorrelated, pure initial states
$\sigma_{kk^{\prime}}$, which we want to study here, the
unperturbed symplectic eigenvalues $\widehat{\nu}^{\,(0)}_{\pm}$ of
$T_{k^{\prime}}\,\sigma_{kk^{\prime}}\,T_{k^{\prime}}$ are degenerate, i.e.,
$\widehat{\nu}^{\,(0)}_{+}=\widehat{\nu}^{\,(0)}_{-}=1$.
This requires the diagonalisation of
the subspaces of the degenerate eigenvalues of the perturbation $\sigma^{c}_{kk^{\prime}}$ to obtain the corrected
smallest symplectic eigenvalue \mbox{$\widehat{\nu}_{-}=1\pm\widehat{\nu}^{\,c}_{-}$}. In other words, the corrections
$\pm\widehat{\nu}^{\,c}_{-}$ to $\,\widehat{\nu}^{\,(0)}_{\pm}=1\,$ are given by the eigenvalues of the matrix
$(\gamma^{c}_{ij})$ with components
\vspace*{-0.2cm}
\begin{align}
    \gamma^{c}_{ij} &=  \left\langle\,e_{i\pm}\,\right|\,
        i\,\Omega\,\widehat{\sigma}^{c}_{kk^{\prime}}\,
        \left|\,e_{j\pm}\,\right\rangle\,,\ \ \ (i,j=1,2)\,,
    \label{eq:projection into eigenspaces}
\end{align}
where $\left|\,e_{j\pm}\,\right\rangle$ is the $j$-th eigenvector of $\,i\,\Omega\,\widehat{\sigma}_{kk^{\prime}}\,$
corresponding to eigenvalue $\pm1$. An important feature of the unperturbed state $\sigma_{kk^{\prime}}$, and
consequently of the eigenvectors of $\,i\,\Omega\,\widehat{\sigma}_{kk^{\prime}}\,$, is its time dependence due to
the free time evolution of the modes, i.e.,
    $\sigma_{kk^{\prime}}\,\rightarrow\,R\,\sigma_{kk^{\prime}}R^{T}\,$.
The local rotation $R$ is represented by an orthogonal, block diagonal matrix $R=\diag\{R_{n}\}$, where the block of
the $n$-th mode can be written as $R_{n}=\Re\,(G_{n})\mathds{1}_{2}+i\,\Im(G_{n})\sigma_{y}$, and $\sigma_{y}$ is the
usual Pauli matrix. The leading order correction to the negativity (\ref{eq:negativity def}) is given
by $|\widehat{\nu}^{\,c}_{-}|/2$. As a particular example we can study the case of symmetric, initial single-mode
squeezing in the modes $k$ and $k^{\prime}$, with (real) squeezing parameters $s_{k}=s_{k^{\prime}}=s$ (see
Eq.~(\ref{eq:single mode squeezing})). In this situation the leading order corrections to the negativity become
\vspace*{-0.2cm}
\begin{align}
    \mathcal{N} &=
    \Bigl( \Re(G^{*}_{k}\beta^{(1)}_{kk^{\prime}})^{2}  +
        (\,\Im(G^{*}_{k}\beta^{(1)}_{kk^{\prime}})\cosh(s)  -  \Im(G^{*}_{k}\alpha^{(1)}_{kk^{\prime}})\sinh(s) )^{2}  \Bigr)^{1/2}
    +O(h^{2})\,.
    \label{eq:single mode squeezed negativity}
\end{align}
For modes of opposite parity, i.e., if $(k+k^{\prime})$ is odd, the linear coefficients are non-zero see Ref.~\cite{BruschiFuentesLouko2012}.
In the case of vanishing squeezing parameters, $s_{k}=s_{k^{\prime}}=0$, the expression
in~(\ref{eq:single mode squeezed negativity}) reduces to $\mathcal{N}=|\beta^{(1)}_{kk^{\prime}}|$,
which is consistent with the expression for the entanglement that is generated from the bosonic
vacuum in the Fock representation, see Ref.~\cite{FriisBruschiLoukoFuentes2012}.
As can easily be seen from Eq.~(\ref{eq:single mode squeezed negativity}) the initial single
mode squeezing introduces a non-negative term into the negativity. Thus the generated negativity is
always enhanced with respect to the vacuum case.
An example for the effect of single-mode squeezing on the generation of entanglement is shown in Fig.~\ref{fig:BBB illustration}(a) for a particular travel scenario of the cavity.
It can be readily observed, that the negativity grows with
$e^{s}$ for $e^{s}\gg e^{-s}$. While this fact can be utilised to enhance the visibility of the
entanglement generation, it also limits the validity of the perturbative regime, which requires
further investigation.

A transparent method to quantify the perturbation of the initial state is the analysis of the
system's mixedness. As discussed in Ref.~\cite{BruschiFuentesLouko2012,BruschiDraganLeeFuentesLouko2012}
the Bogoliubov coefficients that relate the modes $k$ and $k^{\prime}$ can be consistently renormalized
to represent a unitary transformation on the subspace of these two modes. This \emph{two-mode
truncation}, which leaves the linear coefficients $\alpha^{(1)}_{kk^{\prime}}$ and
$\beta^{(1)}_{kk^{\prime}}$ invariant, can be implemented if $k$ and $k^{\prime}$ have opposite
parity. The state of a truncated system of this type will undergo a unitary evolution due to the
cavity motion. In particular, the determinants of the individual mode subspaces, $\,\det C_{kk}\,$
and $\,\det C_{k^{\prime}k^{\prime}}\,$, will be identical to leading order in $h$, i.e., the
state is symmetric.

We then exploit the fact that any pure, symmetric two-mode Gaussian state is locally equivalent
to a two-mode squeezed state~\cite{AdessoIlluminati2005}. This enables us to express the
corresponding two-mode squeezing parameter $r$ as a Maclaurin series in $h$ by using
$|r|=\tfrac{1}{2}\arsinh\sqrt{-\det C_{kk^{\prime}}}$. We recover the expression of
Eq.~(\ref{eq:single mode squeezed negativity}) for the negativity, i.e., $\mathcal{N}=|r|$ as is expected for
two-mode squeezed states. This suggests that Eq.~(\ref{eq:single mode squeezed negativity}) is valid as long
as the perturbation to the mixedness of the system is small, i.e., as long as
$|\det\tilde{\sigma}_{kk^{\prime}}-1|\ll 1$. We compute the determinant of the transformed state for the
symmetrically single-mode squeezed initial state to be
\begin{align}
    \det\tilde{\sigma}_{kk^{\prime}}    &=  1   \,+\,
        4(f^{\beta}_{k\lnot k^{\prime}} + f^{\beta}_{k^{\prime}\lnot k})(\cosh s +1)    \,+\,
        4(f^{\alpha}_{k\lnot k^{\prime}} + f^{\alpha}_{k^{\prime}\lnot k})(\cosh s -1) \nonumber \\[1mm]
        &
        \ \ -\,4\sinh s\,\sum\limits_{n\neq k,k^{\prime}}
        \Re\bigl(\,\alpha^{(1)}_{nk}\beta^{(1)*}_{nk}\,+\,\alpha^{(1)}_{nk^{\prime}}\beta^{(1)*}_{nk^{\prime}}\,\bigr)\,,
    \label{eq:sym squeezed determinant}
\end{align}
where $f^{\alpha}_{k\lnot k^{\prime}}=\tfrac{1}{2}\sum_{n\neq k^{\prime}}|\alpha^{(1)}_{nk}|^{2}$ and
$f^{\beta}_{k\lnot k^{\prime}}=\tfrac{1}{2}\sum_{n\neq k^{\prime}}|\beta^{(1)}_{nk}|^{2}$. The $\alpha$-coefficients
grow with the mode numbers of the selected modes~\cite{BruschiFuentesLouko2012}, while $|\beta^{(1)}_{mn}|\rightarrow0$
if $m\rightarrow\infty$ or $n\rightarrow\infty$. We can thus constrain the range of validity of the perturbative result
of Eq.~(\ref{eq:single mode squeezed negativity}) by the simple requirement
$F_{k,k^{\prime}}(\cosh s -1)\ll1$, where $F_{k,k^{\prime}}=(f^{\alpha}_{k\lnot k^{\prime}} + f^{\alpha}_{k^{\prime}\lnot k})$. Sample plots for $F_{k,k^{\prime}}/h$ are shown in Fig.~\ref{fig:BBB illustration}(b).
Additionally, when a massive, bosonic field is considered the mass of the field excitations must be appropriately restricted,
see Ref.~\cite{BruschiFuentesLouko2012}.
However, as long as these conditions
are satisfied, the single-mode squeezing parameter, the choice of modes and the particular travel scenario for the cavity
can be tuned to enhance the entanglement generation effect.

%

\section{\label{sec:conclusion}Conclusions}

We have presented a general framework that allows the quantification of the entanglement that is generated
between arbitrary modes of bosonic quantum fields by Bogoliubov transformations. Our setup combines techniques from quantum optics, quantum field theory and quantum information procedures to describe the quantum correlations that arise from non-uniform motion, spacetime curvature and quantum optical operations in a covariance matrix formalism. This removes the necessity for cumbersome partial tracing procedures to quantify the entanglement between field modes. For Gaussian initial states we can fully characterise the entanglement that is produced in these situations.

We have discussed a particular example, a cavity containing a relativistic quantum field, in which
non-uniform motion was recently found to create entanglement~\cite{FriisBruschiLoukoFuentes2012}. The Bogoliubov coefficients for this scenario are given as a perturbative expansion and we compute the negativity, the entanglement measure of our choice, to leading order in the expansion parameter. We discuss in detail the regimes in which the perturbative calculations can be trusted.

To leading order in the expansion parameter, the transformed state of two modes of opposite parity is equivalent to a pure, two-mode squeezed state. The corresponding squeezing parameter can be used as an alternative route to quantify the entanglement of the system. We give clear criteria for the validity of
this two-mode truncation of the Bogoliubov transformation. The calculation of the mixedness of the two-mode
state without the truncation provides a simple benchmark against which to judge the limits of the small
parameter expansion.

By application of the methods presented here we are able to extend the previously found results to the important class of Gaussian states. We find that initial single-mode squeezing can enhance the correlations that can be generated from the vacuum, thus greatly improving the prospects of experimental verification of this effect and of its implementation in new quantum technologies.

Under some circumstances, the approach of this article even permits application to particular cases of continuous spectra of quantum fields, such as
a charged, scalar field in an expanding universe~\cite{BallFuentes-SchullerSchuller2006}.

\section*{Acknowledgements}
We thank Gerardo Adesso, David Edward Bruschi, Andrzej Dragan, Lucia Hackerm{\"u}ller, Antony~R.~Lee and Jorma Louko
for helpful discussions and comments.
I.F. and N.F. acknowledge support from EPSRC [CAF Grant No.~EP/G00496X/2 to I.F.].
N.F. thanks the $\chi$-QEN collaboration for support.

\end{document}